\documentclass[aps,prl,superscriptaddress,twocolumn]{revtex4-1}
\usepackage{graphicx}
\usepackage{amsmath}
\usepackage{amssymb}
\usepackage{color}

\newcommand{\gr}{{Gr\"uneisen }}

\begin{document}
\title{Universal Thermodynamic Signature of Self-dual Quantum Critical Points}

\author{Long Zhang}
\email{longzhang@ucas.ac.cn}
\affiliation{Kavli Institute for Theoretical Sciences and CAS Center for Excellence in Topological Quantum Computation, University of Chinese Academy of Sciences, Beijing 100190, China}
\affiliation{Physical Science Laboratory, Huairou National Comprehensive Science Center, Beijing 101400, China}

\date{\today}

\begin{abstract}
Self-duality is an algebraic structure of certain critical theories, which is not encoded in the scaling dimensions and critical exponents. In this work, a universal thermodynamic signature of self-dual quantum critical points (QCPs) is proposed. It is shown that the Gr\"uneisen ratio at a self-dual QCP remains finite as $T\rightarrow 0$, which is in sharp contrast to its universal divergence at a generic QCP without self-duality, $\Gamma(T,g_{c})\sim T^{-1/z\nu}$. This conclusion is drawn based on the hyperscaling theory near the QCP, and has far-reaching implications for experiments and numerical simulations.
\end{abstract}
\maketitle

\emph{Introduction.---}Universality is the hallmark of critical phenomena near continuous phase transitions, which is highly cherished across the entire physics community. The critical points fall into different universality classes according to their critical exponents, which are defined by the singular behaviors of various physical quantities near the critical points. The general theory of critical phenomena has been established since the mid-20th century based on the scaling hypothesis and the renormalization group (RG) theory \cite{Cardy1996}. The critical exponents can be calculated from the scaling dimensions of fields in the critical theory (the eigenvalues of the RG transformation of local operators). It is one of the major theoretical challenges to calculate the critical exponents of different universality classes.

Self-duality is a universal characteristics of certain critical theories, which is not encoded in the critical exponents. It dates back to the Kramers-Wannier duality of the two-dimensional (2D) Ising model \cite{Kramers1941, Kadanoff1971} and was later constructed in a series of models with Abelian symmetries \cite{Wu1976a, Fradkin1978, Zamolodchikov1978, Dotsenko1978, Fradkin1980, Savit1980, Druhl1982}. As a concrete example, in the 1D quantum Ising model with a transverse field (TFIM) \cite{Pfeuty1970},
\begin{equation}
H(g)=-\sum_{i}\sigma_{i}^{x}\sigma_{i+1}^{x}-g\sum_{i}\sigma_{i}^{z},
\end{equation}
the self-duality transformation is defined by \cite{Fradkin1978}
\begin{equation}
U\sigma_{i}^{x}U^{-1}= \prod_{l\leq i}\sigma_{l}^{z},\quad U\sigma_{i}^{z}U^{-1}= \sigma_{i}^{x}\sigma_{i+1}^{x}.
\end{equation}
The Hamiltonian transforms as $UH(g)U^{-1}= gH(1/g)$. The weak-field ($g<1$) and the strong-field ($g>1$) phases are mapped into each other, leaving the quantum critical point (QCP) $g_{c}=1$ unchanged. The spin operator $\sigma_{i}^{x}$ is mapped into the disordering operator $\prod_{l\leq i}\sigma_{l}^{z}$ that creates or annihilates a domain wall, and vice versa. Therefore, the self-duality is a $\mathbb{Z}_{2}$ symmetry acting on fields (operators) in the critical theory at the QCP.

The self-duality provides profound insights into the nature of phases and critical points, e.g., the nonperturbative correspondence between correlations functions in different phases \cite{Kadanoff1971}. However, the experimental evidence of self-duality is rare. In the 2D superconductor-insulator transition \cite{Fisher1989, Fisher1989a, Fisher1990, Fisher1990a} and the quantum Hall-insulator transition \cite{Kivelson1992, Shimshoni1997}, the postulated particle-vortex duality is supported by the universal resistivity at the QCP \cite{Bollinger2011, Breznay2016} and the reflection symmetry of nonlinear $I$-$V$ curves \cite{Shahar1996}. A more universal signature of self-duality remains to be unveiled.

\begin{figure}[b]
\includegraphics[width=0.48\textwidth]{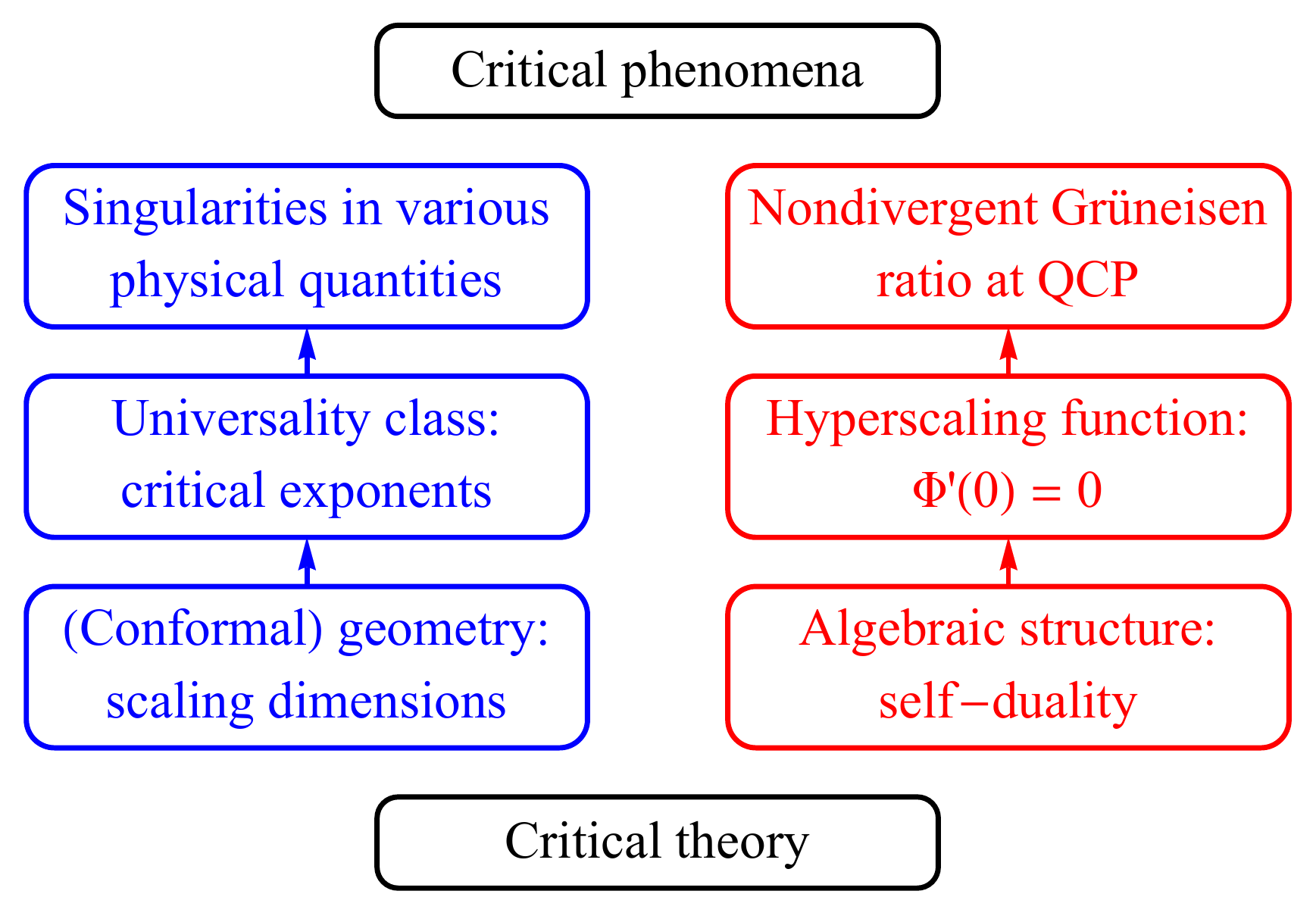}
\caption{The link between the critical theory and critical phenomena. In the general theory of critical phenomena, the universality class is defined based on the critical exponents, which are determined by the scaling dimensions of fields (operators) under the scale (or conformal) transformation. On the other hand, self-duality is an algebraic structure of certain critical theories, which is not encoded in the scaling dimensions. As shown in this work, the self-duality implies that the derivative of the hyperscaling function vanishes at the QCP, and the \gr ratio (GR) does not diverge as $T\rightarrow 0$. This provides a universal thermodynamic signature of self-dual QCPs.}
\label{fig:link}
\end{figure}

In this work, I propose a universal thermodynamic signature of self-dual QCPs. It is shown that the self-duality implies a nondivergent \gr ratio (GR) in the quantum critical regime (QCR) (see Fig. \ref{fig:link}), which is a measurable hallmark of self-dual QCPs with far-reaching implications for experiments.

The GR $\Gamma$ is usually defined as the ratio between the thermal expansion coefficient $\alpha$ and the molar specific heat $c_{p}$,
\begin{equation}
\Gamma=\frac{\alpha}{c_{p}}=\frac{(1/V)(\partial V/\partial T)_{p}}{(T/N)(\partial S/\partial T)_{p}},
\end{equation}
in which $N$ is the mole number making the GR independent of the system size. More generally, the pressure $p$ and the volume $V$ in the definition is replaced by a generalized force $g$ (i.e., a tuning parameter in the Hamiltonian) and its conjugate variable $v_{g}=\partial f(T, g)/\partial g$, in which $f(T,g)$ is the free energy density. For example, near a QCP tuned by the magnetic field $B$, the magnetic GR $\Gamma_{m}$ is defined by $\Gamma_{m}=\frac{(\partial M/\partial T)_{B}}{T(\partial S/\partial T)_{B}}$, in which $M$ is the magnetization in the field direction.

Near a generic QCP $g_{c}$, based on the hyperscaling ansatz of the singular part of the free energy,
\begin{equation} \label{eq:hyper}
f_{s}(T,g)\propto t^{(d+z)/z}\Phi\Big(\frac{r}{t^{1/z\nu}}\Big),
\end{equation}
in which $r=(g-g_{c})/g_{c}$ and $t=T/T_{0}$ are reduced scaling variables, and $T_{0}$ is a nonuniversal characteristic temperature scale. $\Phi(x)$ is the smooth hyperscaling function. The GR $\Gamma(T,g)$ is shown to diverge in a universal manner in the QCR \cite{Zhu2003},
\begin{equation} \label{eq:tdiv}
\Gamma(T,g) \sim -G_{T}T^{-1/z\nu},\quad |r|/t^{1/z\nu}\ll 1,
\end{equation}
and changes sign as $g$ is tuned through the QCP at low temperature \cite{Zhu2003, Garst2005},
\begin{equation} \label{eq:gdiv}
\Gamma(T,g)\sim \frac{1}{g-g_{c}},\quad |r|/t^{1/z\nu}\gg 1,
\end{equation}
in which $z$ and $\nu$ are the dynamical and the correlation-length critical exponents, respectively. The prefactor $G_{T}$ is further discussed below. The universal behavior of $\Gamma(T,g)$ has been confirmed near the QCPs of heavy fermion materials \cite{Kuchler2003}, and adopted as a standard thermodynamic probe to quantum criticality \cite{Gegenwart2008, Gegenwart2016}.

Given the success of the above universal scaling form, it is a surprise that $\Gamma(T,g)$ does not diverge as $T\rightarrow 0$ at the QCP of the 1DTFIM. With the exact solution of the 1DTFIM, $\Gamma(T,g)$ at the QCP is found to be precisely $1/2$ at any temperature \cite{Wu2018}. At first sight, this contradicts the scaling form in Eq. (\ref{eq:tdiv}). This is resolved by noting that the prefactor $G_{T}$ is proportional to the derivative of the hyperscaling function at the QCP \cite{Zhu2003},
\begin{equation}
G_{T}=\frac{(d+z-1/\nu)z\Phi'(0)}{d(d+z)\Phi(0)}\frac{T_{0}^{1/z\nu}}{g_{c}},
\end{equation}
and it is $\Phi'(0)$ that vanishes in the 1DTFIM \cite{Garst2005, Wu2018}, which was taken as a characteristic feature of the 1DTFIM \cite{Wu2018}. The nondivergence of $\Gamma(T,g)$ in the QCR was confirmed in the magnetocaloric measurement of an Ising-like spin-$1/2$ antiferromagnetic chain material BaCo$_{2}$V$_{2}$O$_{8}$ \cite{Wang2018a}.

However, as I will show in this work, both the vanishing $\Phi'(0)$ and the nondivergent $\Gamma(T,g)$ in the QCR turn out to be a generic consequence of any self-dual QCPs, among which the 1DTFIM is the simplest example \footnote{After completing this work, I learned that the relation between the vanishing $G_{T}$ and the self-duality in the 1DTFIM was briefly noted in Ref. \cite{Garst2005}. I thank J. Wu for pointing this out to me.}. Therefore, the nondivergence of $\Gamma(T,g)$ is a universal thermodynamic signature of self-duality. I will first give a simple and rigorous proof for QCPs with exact self-duality, and make a heuristic generalization based on the hyperscaling ansatz and the formal RG theory, and then present a few examples. Finally, I will discuss its implications to experiments and possible extensions.

\emph{QCP with exact self-duality.---}Let us first consider a generic QCP with an exact self-duality. Assume a Hamiltonian parametrized by $g$ with a QCP at $g_{c}=1$, $H(g)=H_{1}+gH_{2}$. A unitary or an antiunitary self-duality transformation $U$ satisfies
\begin{equation}
UH_{1}U^{-1}=H_{2},\quad UH_{2}U^{-1}=H_{1}.
\end{equation}
The Hamiltonian at the QCP is invariant under the self-duality transformation, $UH(g_{c})U^{-1}=H(g_{c})$. This definition is not as restricted as it appears. The 1DTFIM and the 1D quantum clock model are two examples among a broad range of exactly self-dual models with Abelian symmetries \cite{Savit1980}.

Taking the thermodynamic expectation values at a finite temperature $T$ at the QCP, the self-duality symmetry guarantees that
\begin{equation}
\langle H_{1}\rangle_{T}=\langle UH_{1}U^{-1}\rangle_{T}=\langle H_{2}\rangle_{T}=\frac{1}{2}\langle H(g_{c})\rangle_{T}.
\end{equation}
Therefore, the GR at the QCP
\begin{equation}
\Gamma(T,g_{c})=\frac{\partial \langle H_{2}\rangle_{T}/\partial T}{\partial \langle H(g_{c})\rangle_{T}/\partial T}=\frac{1}{2},
\end{equation}
which is independent of the temperature. Thereby this is a direct consequence of the exact self-duality without referring to any specific properties of the model.

\emph{Self-duality and hyperscaling.---}Let us give the following heuristic definition of self-duality for more generic QCPs. Suppose that the QCP is controlled by a fixed point $H^{*}$ in the space of all possible couplings (theory space), and the tuning parameter $g-g_{c}$ couples to the relevant operator $R$, which has an RG eigenvalue $y=1/\nu>0$. The irrelevant operators are denoted by $I_{n}$ with RG eigenvalues $y_{n}<0$. The self-duality $U$ is a unitary or an antiunitary transformation on the operators,
\begin{subequations}
\begin{align}
UH^{*}U^{-1}& = H^{*}+\sum_{n}a_{n}I_{n}, 	\\
URU^{-1}	& = -R + \sum_{n}b_{n}I_{n},	\\
UI_{n}U^{-1}& = \sum_{m}u_{nm}I_{m}.
\end{align}
\end{subequations}
In other words, the self-duality is a $\mathbb{Z}_{2}$ transformation in the theory space up to irrelevant operators. It may also act nontrivially on other operators that do not directly appear in the Hamiltonian, e.g., it interchanges the spin and the disordering operators in the 1DTFIM.

Expand the Hamiltonian around the fixed point,
\begin{equation}
H(g)=H^{*}+\delta g R+\sum_{n}\alpha_{n}(g)I_{n},
\end{equation}
where $\delta g=g-g_{c}$. Based on the formal RG theory, the singular part of the free energy density can be expressed by \cite{Cardy1996}
\begin{equation}
\begin{split}
	& f_{s}(T,g,\{\alpha_{n}(g)\}) \\
= 	& t^{(d+z)/z}\Phi\Big(\frac{r}{t^{1/z\nu}},\{\alpha_{n}(g)t^{|y_{n}|/z}\}\Big)	\\
=	& t^{(d+z)/z}\Phi\Big(\frac{r}{t^{1/z\nu}}\Big)+\ldots,
\end{split}
\end{equation}
in which $\Phi(x)$ is the universal hyperscaling function appearing in Eq. (\ref{eq:hyper}), and the ellipsis denotes the subleading corrections contributed by the irrelevant terms. The dynamical exponent $z$ enters because the temperature $T$ is also rescaled in the RG transformation.

The self-duality then implies that
\begin{equation}
\begin{split}
e^{-Nf(T,g)/T} & = \mathrm{Tr}e^{-H(g)/T} 	\\
& = \mathrm{Tr}\big(Ue^{-H(g)/T}U^{-1}\big)	\\
& = \mathrm{Tr}e^{-(H^{*}-\delta g R+\sum_{n}\beta_{n}(g)I_{n})/T},
\end{split}
\end{equation}
in which $\beta_{n}(g)=a_{n}+\delta g b_{n}+\sum_{m}\alpha_{m}(g)u_{mn}$. Therefore, the singular part of the free energy satisfies
\begin{equation}
f_{s}(T,g_{c}+\delta g,\{\alpha_{n}(g)\})=f_{s}(T,g_{c}-\delta g,\{\beta_{n}(g)\}),
\end{equation}
which implies
\begin{equation}
\Phi\Big(\frac{r}{t^{1/z\nu}}\Big)=\Phi\Big(-\frac{r}{t^{1/z\nu}}\Big).
\end{equation}
Therefore, the self-duality of the QCP implies $\Phi'(0)=0$ and thereby the saturation of $\Gamma(T,g)$ as the temperature decreases in the QCR, where $|r|/t^{1/z\nu}\ll 1$, and particularly its nondivergence at the QCP.

\begin{figure}[bt]
\includegraphics[width=0.48\textwidth]{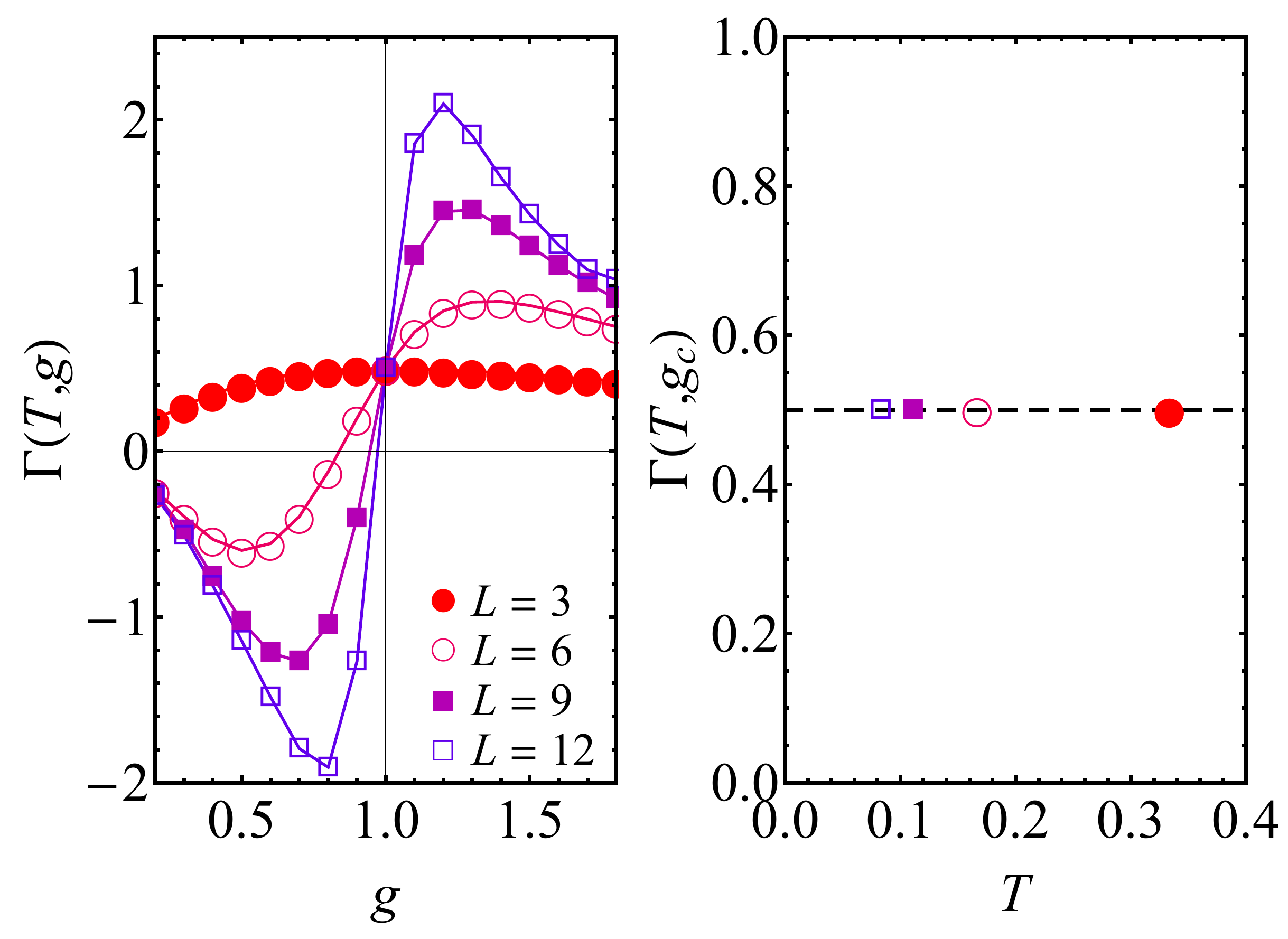}
\caption{Left: the GR of the Ising domain wall model near its QCP $g_{c}=1$ on finite-size lattices with periodic boundary condition. The temperature $T=1/L$. Right: $\Gamma(T,g_{c})=1/2$ for any lattice sizes and temperature.}
\label{fig:dw}
\end{figure}

\emph{Example 1: Ising domain wall model.---}Consider the following 1D spin-1/2 model,
\begin{equation}
H(g)=-\sum_{i}\sigma_{i}^{z}\sigma_{i+1}^{x}\sigma_{i+2}^{z}-g\sum_{i}\sigma_{i}^{y}.
\end{equation}
It hosts an exact self-duality \cite{Bao2018},
\begin{equation}
U=\prod_{i}\sigma_{i}^{x}\prod_{i}e^{-(i\pi/4)(\sigma_{i}^{z}\sigma_{i+1}^{z}-\sigma_{i}^{z}-1)},
\end{equation}
which satisfies $UH(g)U^{-1}=gH(1/g)$. There is a self-dual QCP at $g_{c}=1$. It was derived as the effective model of the domain wall between the toric code and the double semion topological order states, and dubbed the Ising domain wall model (IDWM) \cite{Bao2018}. The self-duality of the IDWM is a local transformation, thus it is exact even on a finite-size lattice with periodic boundary condition (p.b.c.), which is in sharp contrast to the nonlocal self-duality of the 1DTFIM and related models.

The IDWM is numerically diagonalized on finite lattices with p.b.c. The GR near the QCP is shown in Fig. \ref{fig:dw}. While the sign change of $\Gamma(T,g)$ at low temperature near the QCP is consistent with the general scaling form in Eq. (\ref{eq:gdiv}), the curves intersect precisely at the QCP, and $\Gamma(T,g_{c})=1/2$ for all temperature (Fig. \ref{fig:dw}, right panel). As shown above, this is a thermodynamic signature of the exactly self-dual QCP.

\emph{Example 2: 2D topological transition.---}The following lattice model of 2D free fermions \cite{Bernevig2006a} describes the topological quantum phase transition (QPT) from a quantum anomalous Hall insulator ($m<0$) to a normal insulator ($m>0$),
\begin{subequations} \label{eq:bhz}
\begin{gather}
H	=\sum_{k}\psi_{k}^{\dag}h(k)\psi_{k},\quad \psi_{k}=
\begin{pmatrix}
c_{k}	\\
v_{k}
\end{pmatrix},\\
h(k)=
\begin{pmatrix}
\epsilon_{c}(k)+m 	& \Delta(k) \\
\Delta(k)^{*}		& \epsilon_{v}(k)-m
\end{pmatrix},
\end{gather}
\end{subequations}
in which $\epsilon_{c}(k)=-2t_{c}(\cos k_{x}+\cos k_{y}-2)$, $\epsilon_{v}(k)=2t_{v}(\cos k_{x}+\cos k_{y}-2)$, and $\Delta(k)=\Delta(\sin k_{x}-i\sin k_{y})$. Close to the QCP at $m=0$, the low-energy physics is captured by the massive Dirac Hamiltonian,
\begin{equation}
h(k) \simeq \Delta (k_{x}\sigma_{x}+ k_{y}\sigma_{y}) +m\sigma_{z}.
\end{equation}
The antiunitary transformation $U\psi_{k}U^{-1}=\mathcal{K}\sigma_{x}\psi_{k}$ ($\mathcal{K}$ is the complex conjugate) changes the sign of $m$ and, thus, interchanges the two phases and leaves the QCP unchanged up to irrelevant terms. Therefore, it serves as a self-duality of this topological QPT.

\begin{figure}[bt]
\includegraphics[width=0.48\textwidth]{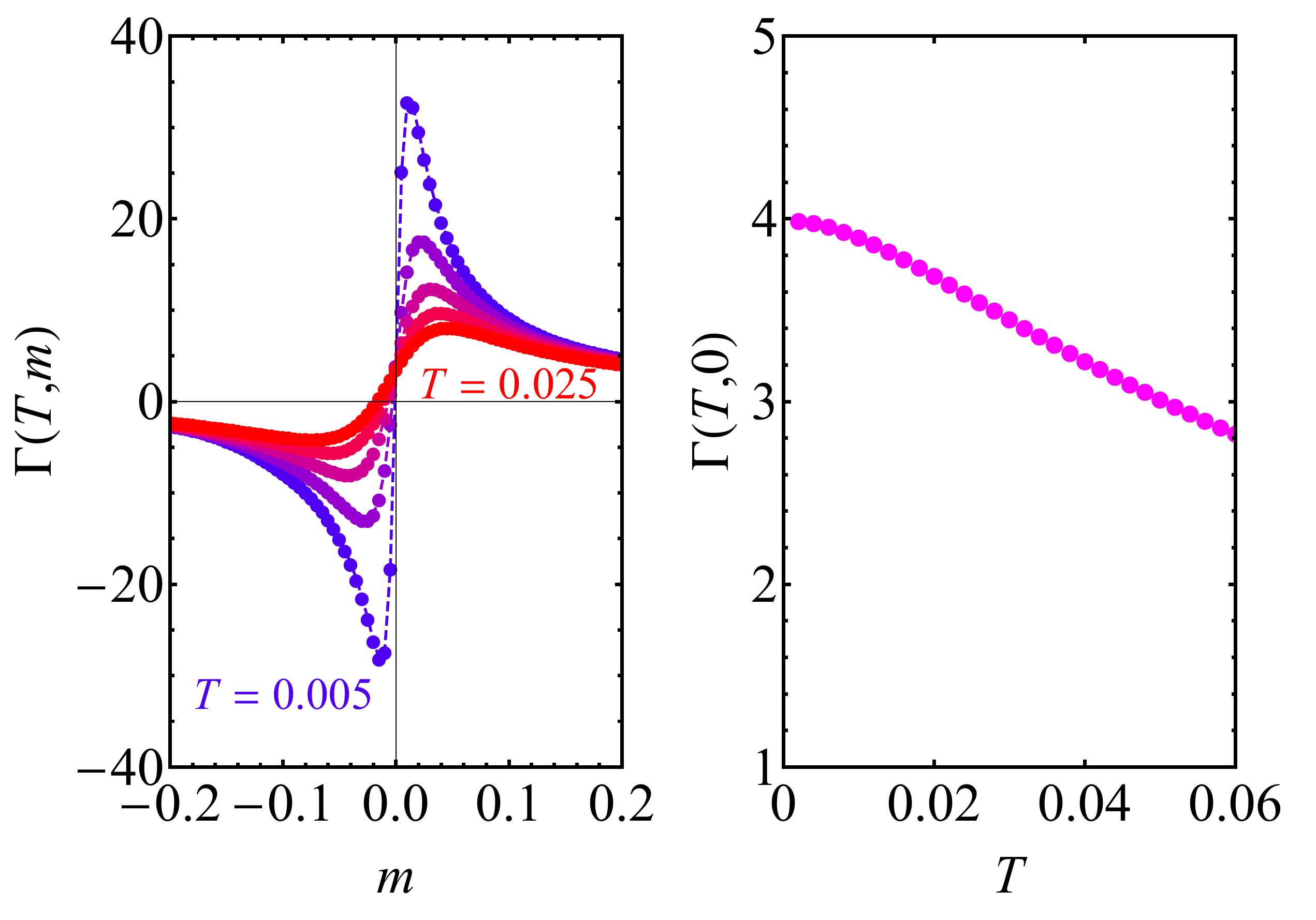}
\caption{Left: the GR near the 2D topological quantum phase transition described by Eq. (\ref{eq:bhz}). Parameters: $t_{c}=1.2$, $t_{v}=0.8$, and $\Delta=0.5$. Right: at the QCP, $\Gamma(T,0)$ saturates as $T\rightarrow 0$.}
\label{fig:bhz}
\end{figure}

The GR $\Gamma(T,m)$ near the QCP is shown in Fig. \ref{fig:bhz}. At low temperature, its sign change near the QCP is again captured by the general scaling form, Eq. (\ref{eq:gdiv}). However, at the QCP, $\Gamma(T,0)$ saturates as $T\rightarrow 0$ (Fig. \ref{fig:bhz}, right panel), exhibiting the thermodynamic signature of self-duality.

\emph{Example 3: QCP without self-duality.---}In order to highlight the peculiarity of the self-dual QCPs studied above, let us contrast them with a QCP without self-duality. The chemical potential-tuned metal-insulator transition (MIT) in 1D is described by
\begin{equation} \label{eq:sm}
H=\sum_{k}\xi_{k}c_{k}^{\dag}c_{k}, \quad \xi_{k}=-2\cos k-\mu.
\end{equation}
The system is metallic with a partially filled band for $|\mu|<2$, and insulating with an empty conduction band for $\mu<-2$. The MIT takes place at $\mu _{c}=-2$. The critical exponents $z=2$ and $\nu=1/2$ can be extracted from the single-particle Green's function near the QCP. This QCP cannot be self-dual because the energy spectra are gapless and gapped in the two phases, respectively; thus they cannot be mapped into each other by a unitary or an antiunitary transformation.

\begin{figure}[bt]
\includegraphics[width=0.48\textwidth]{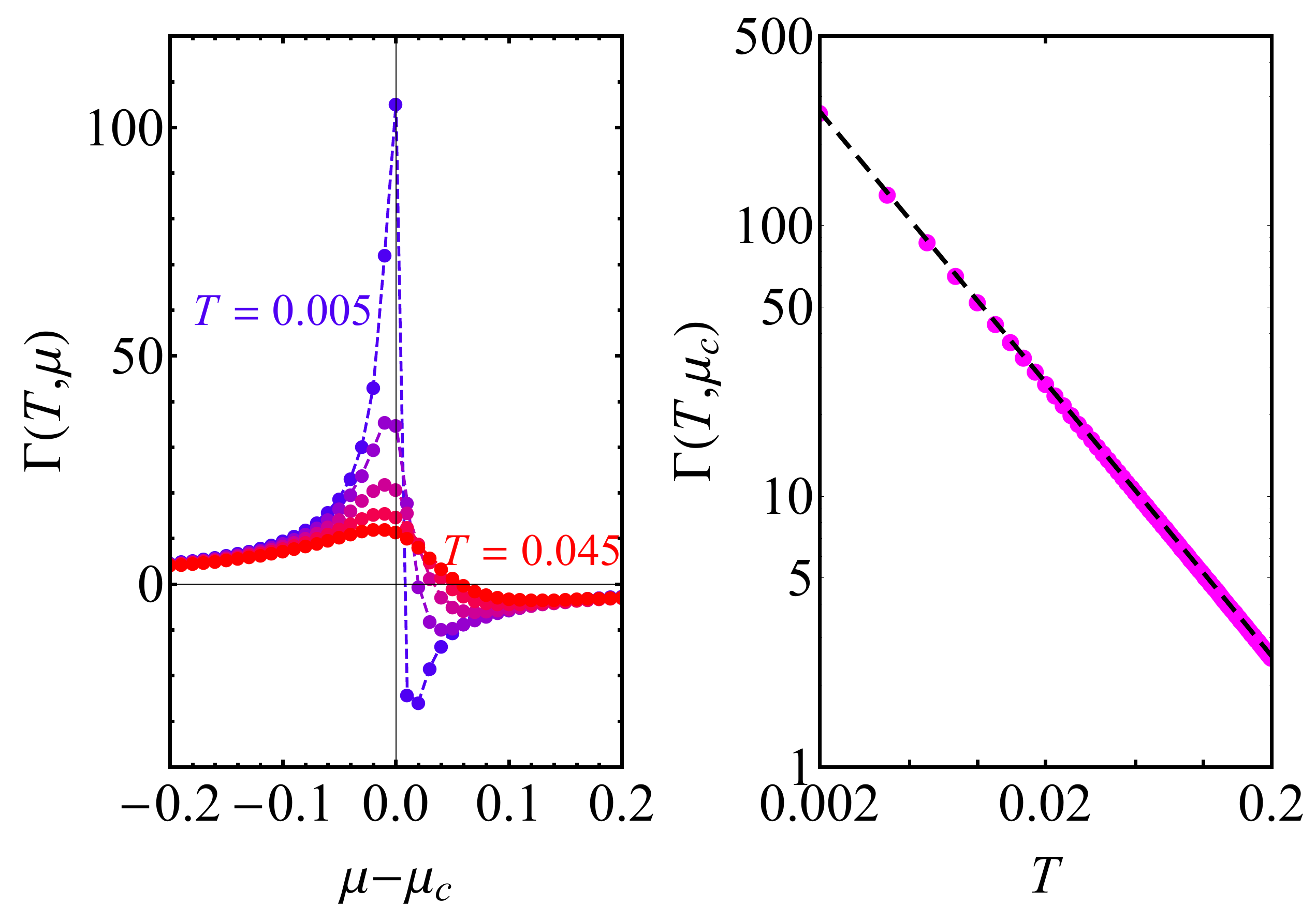}
\caption{Left: the GR near the 1D chemical potential-tuned metal-insulator transition described by Eq. (\ref{eq:sm}). Right: at the QCP, $\Gamma(T,\mu_c)$ diverges as $T\rightarrow 0$. The power-law fitting yields $\Gamma(T,\mu_c)=-0.067+0.527/T$ (the dashed line).}
\label{fig:sm}
\end{figure}

The GR near the QCP is shown in Fig. \ref{fig:sm}. In the low-temperature regime, $\Gamma(T,\mu)$ changes sign near the QCP as the scaling form of Eq. (\ref{eq:gdiv}) like other QCPs. However, at the QCP, $\Gamma(T,\mu_c)$ also diverges as $T\rightarrow 0$, which is in sharp contrast to the self-dual QCPs. The power-law fitting (Fig. \ref{fig:sm}, right panel) yields $\Gamma(T,\mu_c)\sim T^{-1}$, fully consistent with the general scaling form in Eq. (\ref{eq:tdiv}).

\emph{Discussion: Nondivergent GR implies self-duality.---}Is self-duality a necessary condition for the nondivergence of GR at the QCP? I will argue that this is true by explicitly constructing a unitary self-duality transformation near the QCP.

Suppose that the QCP is perturbed by a relevant operator $R$, $H(g)=H(g_{c})+\delta g R$, and the GR $\Gamma(T, g_{c})$ is finite as $T\rightarrow 0$ and can be expanded as $\Gamma(T,g_{c})= \Gamma(0,g_{c})+\sum_{n=1}^{\infty}\frac{1}{n!}\Gamma^{(n)}(0,g_{c})T^{n}$. In the eigenstate basis of $H(g_{c})$,
\begin{equation}
H(g_{c})=\int_{0}^{\infty} d\epsilon D(\epsilon,g_{c})\epsilon |\epsilon(g_{c})\rangle \langle \epsilon(g_{c})|,
\end{equation}
in which $D(\epsilon,g_{c})$ is the density of states of $H(g_{c})$ and $|\epsilon(g_{c})\rangle$ is the eigenstate of $H(g_{c})$ with energy $\epsilon$ \footnote{The possible degeneracy can be lifted by irrelevant perturbations.}. The deformed Hamiltonian $\tilde{H}(g)=H(g)-\delta g(\Gamma(0,g_{c})H(g_{c})+\delta H)$ satisfies
\begin{equation}
\tilde{\Gamma}(T,g_{c} )=\frac{\partial\langle R -\Gamma(0,g_{c})H(g_{c}) -\delta H \rangle /\partial T}{\partial\langle H(g_{c})\rangle /\partial T}=0
\end{equation}
at any $T$, in which
\begin{align}
\delta H &= \int_{0}^{\infty}d\epsilon \delta D(\epsilon)\epsilon |\epsilon (g_{c} )\rangle\langle\epsilon (g_{c})|, \\
\delta D(\epsilon) &= \sum_{n=1}^\infty \frac{\Gamma^{(n)}(0,g_{c})}{n!(n-1)!\epsilon^{2}} \int_0^\epsilon d\omega (\epsilon -\omega)^{n-1} D(\omega,g_{c}) \omega^{2}.
\end{align}
Furthermore, given that $H(g_{c})$ is strictly marginal at the QCP, $\delta H$ is irrelevant from dimensional analysis.

The vanishing of $\tilde{\Gamma}(T,g_{c})$ at any $T$ implies that the deformed entropy $\tilde{S}(T,g)$ of $\tilde{H}(g)$ satisfies $\partial \tilde{S}(T,g_{c})/\partial g=0$ at any $T$; thus the deformed density of states $\tilde{D}(\epsilon, g)$ satisfies $\partial \tilde{D}(\epsilon,g_{c})/\partial g= 0$ for any $\epsilon$. Therefore,
\begin{equation}
\tilde{D}(\epsilon, g_{c}+\delta g)=\tilde{D}(\epsilon, g_{c}-\delta g)+O(\delta g^{3}).
\end{equation}
In other words, there is a one-to-one correspondence between the spectra of $\tilde{H}(g_{c}+\delta g)$ and $\tilde{H}(g_{c}-\delta g)$ up to vanishingly small energy level shift.

Denoting the eigenstates of $\tilde{H}(g_{c}\pm\delta g)$ by $|\tilde{\epsilon}(g_{c}\pm \delta g)\rangle$, a unitary transformation $U(\delta g)$ can be constructed for each $\delta g$,
\begin{equation}
U(\delta g)|\tilde{\epsilon}(g_{c}+\delta g)\rangle =|\tilde{\epsilon}(g_{c}-\delta g)\rangle,
\end{equation}
under which $\tilde{H}(g)$ transforms as
\begin{equation}
U(\delta g)\tilde{H}(g_{c}+\delta g)U(\delta g)^{-1}=\tilde{H}(g_{c}-\delta g)+O(\delta g^{3}).
\end{equation}
$U(\delta g)$ converges to a unitary transformation $U$ as $\delta g\rightarrow 0$, which satisfies
\begin{align}
UH(g_{c})U^{-1} &= H(g_{c}) + O(\delta g^{2}),\\
U(R-\delta H)U^{-1} &= -R+2\Gamma(0,g_{c})H(g_{c})+\delta H +O(\delta g^{2}),
\end{align}
thus is not an identity operator. The original Hamiltonian $H(g)$ transforms as
\begin{equation}
UH(g_{c}+\delta g)U^{-1}\simeq (1+2\Gamma(0,g_{c})\delta g)H(g_{c}-\delta g),
\end{equation}
where irrelevant operators are omitted. This has a similar form as the exact self-duality up to numerical prefactors and irrelevant operators, both of which only modify the nonuniversal behavior of the GR quantitatively. Therefore, $U$ is the desired unitary self-duality transformation near the QCP.

\emph{Summary and outlook.---}In summary, it is shown that the GR does not diverge as $T\rightarrow 0$ in the QCR of a self-dual QCP. This is in sharp contrast to its universally diverging form at a generic QCP without self-duality. Therefore, it serves as a universal thermodynamic signature of self-dual QCPs.

The self-duality is an algebraic structure of the critical theory, and is not encoded in the scaling dimensions and the critical exponents. Indeed, self-duality has been constructed in various critical points with distinct critical exponents. In this work, it is shown to be reflected in the hyperscaling function $\Phi(x)$: $\Phi'(0)=0$ at a self-dual QCP. This inspires us to further unveil the rich information encoded in the hyperscaling functions with particular attention on the algebraic structure of the critical theory (see Fig. \ref{fig:link}).

This work has far-reaching implications for experiments. Let us first note that the nondivergent GR was observed in the magnetocaloric measurements on BaCo$_{2}$V$_{2}$O$_{8}$, but taken as a special feature of the 1DTFIM \cite{Wang2018a}. As shown in this work, this is indeed the first thermodynamic evidence of a self-dual QCP. It is greatly desirable to perform similar magnetocaloric measurements on the 2D superconductor-insulator transition and the quantum Hall-insulator transition, both of which have been postulated as self-dual QCPs \cite{Fisher1990, Fisher1990a, Shimshoni1997}.

Moreover, this thermodynamic signature may be adopted in numerical simulations to test the postulated self-dual QCPs. By comparing the low-temperature behavior of the GR at the QCP with the general diverging form, $\Gamma(T,g_{c})\sim T^{-1/z\nu}$ ($z$ and $\nu$ can be extracted from other physical quantities), one may provide a thermodynamic evidence to or falsify the postulated self-duality. In order to put this into work, it is necessary to reach a better understanding of the finite-size and the boundary effects on QCPs with nonlocal self-duality. This will be presented in a separate work \cite{Zhang2019a}.

\acknowledgments
I am grateful to C. Ding, D. N. Sheng, J. Wu, and F. C. Zhang for stimulating discussions. This work is supported by National Key R\&D Program of China (No. 2018YFA0305800), National Natural Science Foundation of China (No. 11804337), Strategic Priority Research Program of CAS (No. XDB28000000), and Beijing Municipal Science and Technology Commission (No. Z181100004218001).

\bibliography{/Dropbox/ResearchNotes/Bibtex/library,/Dropbox/ResearchNotes/Bibtex/Books}
\end{document}